\documentstyle[aps,manuscript]{revtex}

\tightenlines

\begin{document}
\newcommand{\lc}{\left[}
\newcommand{\rc}{\right]}
\newcommand{\lex}{\left<}
\newcommand{\rex}{\right>}
\newcommand{\ex}[1]{\left<#1\right>}
\newcommand{\ket}[1]{\left|#1\right>}
\newcommand{\bra}[1]{\left<#1\right|}
\newcommand{\J}{\hat{J}}
\title{Weak force detection using a double Bose-Einstein condensate}
\author{J.\ F.\ Corney$^1$, G.\ J.\ Milburn$^1$ and Weiping Zhang$^2$} 
\address{$^1$Department of Physics, The University of Queensland, QLD 4072,
Australia}
\address{$^2$School of Mathematics, Physics, Computing and Electronics, 
Macquarie University, Sydney, NSW 2109, Australia}
\date{\today}
\maketitle

\begin{abstract}

A Bose-Einstein condensate may be used to make precise measurements of weak forces, utilizing the macroscopic occupation of a single quantum state.  We present a scheme which uses a condensate in a double well potential to do this.  The required initial state of the condensate is discussed, and the limitations on the sensitivity due to atom collisions and external coupling are analyzed.

\end{abstract}

\newpage

\section{Introduction}
The experimental demonstration of Bose-Einstein condensation (BEC) in trapped quantum gases of alkali atoms\cite{Anderson95,Hulet95,Ketterle95,BEC97} opens up great opportunities in atomic physics, condensed matter physics and quantum optics. As a macroscopic quantum object, the physical properties of the Bose-Einstein condensate (BEC)  have naturally attracted great attention. While much experimental and theoretical work has already been done to investigate the basic properties of the condensates, possible  practical application is still an open question. Very recently, experimentalists at JILA have attempted to build an atomic clock, based on persistent tunneling oscillations between two coupled BECs\cite{ScienceNews,HalMatetal98,HalMatWieCor98}. Although this type of atomic clock is still crude, the experiments have stimulated the search for practical applications of an atomic BEC.

In this paper, we propose a scheme to detect weak forces by
employing a BEC confined in a double-well potential. The basic idea for
this purpose is shown in fig.(\ref{fig1}). The condensate is prepared
initially in
a coherent superposition of the extremal eigenstates of the operator
of particle number difference between the two wells. The requirement for
such a state will be discussed below, but it should be noted that this
state is an entangled state and cannot be described
by semiclassical mean field theory.   Under the action of a weak-force
for a time $\tau$, the condensate will experience a phase shift. The phase
shift can be detected by using the technique analogous to the Ramsey
interference. The interference fringes can be read out by performing
a homodyne measurement of the optical phase shift due to the dispersive
interaction of the condensate localized in one
well and an optical field mode.  We have analyzed the limitations on the
accuracy of the scheme and show
that a high-precision measurement can be achieved if the condensate
contains a large number of coherently condensed atoms.
This result is due to the quantum entanglement\cite{HueMacetal97} inherent
in the initial superposition state of BEC.
 
\section{Two mode system}

We consider the case where a condensate has formed in a quartic double-well potential: 
\begin{equation} 
V(x)=b\left(x^2-x_0^2\right)^2,
\end{equation}
which has minima at $x = \pm x_0$ and a trap frequency of $\omega_0 = \sqrt{8b/m}x_0$.  In a two mode approximation where the total atom number $\hat{N} = N$ is conserved, the system hamiltonian can be described in terms of angular momentum operators\cite{MilCorWriWal97,CorMil98}:
\begin{equation}
\hat{H}= \hbar\Omega\J_z+2\hbar \kappa\J_x^2. \label{ham}
\end{equation}
The commutation relations for these operators are
\begin{equation}
\lc \J_i,\J_j\rc = i\epsilon_{ijk}\J_k 
\end{equation}
where $\epsilon_{ijk}$ allows cyclic permutations of $i,j,k \in\{x,y,z\}$.
 
The operator $\J_x$ gives the condensate particle number difference between the two wells,  $\J_y$ corresponds to the momentum induced by tunneling, and $\J_z$ is the difference in occupation between the upper and lower energy eigenstates of the potential.  The splitting between these levels is given by $\Omega$, which is the tunneling frequency, and $\kappa$ corresponds to the strength of the interparticle hard-sphere interactions.  

The two mode approximation is valid in the regime where the overlap between the single particle ground state modes of each well is small and where the many-body effects do not affect significantly the properties of these modes.  These conditions lead to\cite{MilCorWriWal97,CorMil98}
\begin{equation}
\frac{\Omega}{\omega_0} <<1, \hspace{1cm} N<<\frac{r_0}{|a|},
\end{equation}
where $a$ is the scattering length which determines the strength of the two-atom collisions and where $r_0=\sqrt{\hbar /2m\omega_0}$ characterizes the size of the wells.

\section{Measurement Scheme}

As is shown in fig.(\ref{fig1}), the detection of the weak force proceeds in several 
stages. The first step is to prepare the initial state of the system in a
quantum state which can optimize the precision of the measurement by 
controlling the atomic parameters. With this in mind, we consider the
weak tunneling limit $\Omega << \kappa N$. In this case, the ground state 
of the hamiltonian (eq.(\ref{ham})) for attractive interactions ($\kappa < 0$) is 
a superposition of the extremal eigenstates of the operator $\J_x$\cite{HarWisMil98,SteCol98,CirLewMolZol98}:
\begin{equation}
\ket{E} = \frac{1}{\sqrt{2}} \left( \ket{j,-j}_x + \ket{j,j}_x\right).
\label{gnd}
\end{equation}
We will see that with such an initial state the precision of the 
measurement can be controlled by the total number of atoms in the
condensate. 
This state is easily represented on the Bloch sphere (fig.(\ref{fig1}(a))) by two diametrically opposed points on the equator.  In principle, such a state could be prepared 
by allowing the atoms to condense into the ground state of a double-well potential.  By increasing the height or width of the barrier, 
the tunneling rate $\Omega$ could be adiabatically decreased\cite{CirLewMolZol98} until the ground state evolves into the superposition state (eq.(\ref{gnd})).

The second step in the measurement process, after having achieved the 
ground state described by eq.(\ref{gnd}), and before the tunneling is 
turned on, is to allow the weak force to act for a certain time $\tau$.  
This force may be due to a varying gravitational field, for example, and has 
the effect of adding a linear ramp to the potential:
\begin{equation}
\hat{H_F}= \hbar\Delta\J_x,
\end{equation}
where $\Delta$ is the frequency shift induced by the weak force 
(see fig.(\ref{fig1}(b))) and $j=N/2$.
To avoid the self-phase shift due to interatomic collisions
(nonlinear term in the hamiltonian eq.(\ref{ham})), a technique such as that using Feshbach resonances\cite{InoAndetal98}
may be used to tune the size of the interatomic interaction close to zero at this stage. 
As a result, the initial ground state (eq.(\ref{gnd})) evolves into the
superposition under the weak force: 
\begin{equation}
\ket{\Psi(\tau)} = \frac{1}{\sqrt{2}}
\left(e^{i\Delta j \tau}\ket{j,-j}_x + e^{-i\Delta j \tau}\ket{j,j}_x\right)
\label{super}
\end{equation}
The state now contains the phase-shift induced by the weak force. This phase shift is ``amplified'' $2j$ times the single-particle value
due to the specially prepared initial state.

In the next step we
propose to use a technique analogous to Ramsey interferometry.  In order to detect the induced phase shift, we rotate 
the state (eq.(\ref{super})) around the Bloch sphere by $90^0$. Such a rotation can
be achieved by turning on the tunneling between the two wells of the trap for a time $t_{\pi/2} = \frac{\pi}{2 \Omega}$.
If the interatomic interaction is strong, the rotation operation will
be affected by the nonlinear term in eq.(\ref{ham}), which will distort
the final state from the y-axis on the Bloch sphere and affect the 
precision of the measurement. To avoid this, we continue to use the Feshbach resonance to suppress the collisions.  The state is then rotated to 
\begin{equation}
\ket{\Psi(\tau+t_{\pi/2})} = \frac{1}{\sqrt{2}}\left(e^{i\Delta j \tau}\ket{j,-j}_y + e^{-i\Delta j \tau}\ket{j,j}_y\right).
\end{equation}

Finally, a number measurement can be performed on one of the condensates.  This corresponds to a projection onto the $\J_x$ eigenstates.  The resultant probability distribution is 
\begin{equation}
P_x(m) = \frac{1}{2}\left|e^{i\Delta \tau j} {}_x\hspace{-2pt}\ex{j,m|j,-j}_y + e^{-i\Delta \tau j} {}_x\hspace{-2pt}\ex{j,m|j,j}_y\right|^2.
\end{equation}
Now the inner product of the $J_x$ and $J_y$ eigenstates will be a binomial function of $m$ peaked around $m = 0$, with 
\begin{equation}
{}_x\hspace{-2pt}\ex{j,m|j,-j}_y = e^{-im\pi}{}_x\hspace{-2pt}\ex{j,m|j,j}_y.
\end{equation}
This leads to 
\begin{eqnarray}
P_x(m) &=& 2 \cos^2(\Delta j\tau + m\pi/2)\left| {}_x\hspace{-2pt} \ex{j,m|j,j}_y \right|^2 \\ \label{prob}
 &=&  \left\{ \begin{array}{l}
 2 \cos^2{\Delta j\tau}\left| {}_x\hspace{-2pt} \ex{j,m|j,j}_y  \right|^2 \; m \: {\rm even} \\
 2 \sin^2{\Delta j\tau}\left| {}_x\hspace{-2pt} \ex{j,m|j,j}_y \right|^2 \;m \: {\rm odd} 
\end{array}
\right. ,
\end{eqnarray}
which describes how the output fringes are shifted by the presence of the force $\Delta$.  In the absence of the force, all the odd fringes are absent, but for $\Delta \neq 0$, the probability that a particular measurement will fall on an odd fringe is
\begin{equation}
Pr({\rm odd}) = \sin^2{\Delta j\tau} \simeq (\Delta j \tau)^2 
\label{odd}
\end{equation}
for small $\Delta$.

\section{Measurement readout}

The measurement of atom number is effected through a homodyne scheme\cite{CorMil98}.  The condensate is placed in an optical cavity, which at the time of the readout stage of the measurement contains a light field which is highly driven and damped.  The optical field is thus in a coherent state with amplitude $\alpha_0$.  The light field is detuned from any atomic resonance and so the condensate merely imposes a phase shift on the light.  This can be detected by measuring the quadrature components of the field.  For a dispersive interaction which acts on a timescale $t_I$ over which the dynamics of the condensate itself is negligible,
\begin{equation}
\hat{H_I}= \chi\J_x a^\dagger a,
\end{equation}
where $\chi$ is the measurement strength and $a^\dagger$, $a$ are the light field creation and annihilation operators.  
The quadrature components $\hat{X} = a^\dagger+ a$ and $\hat{Y} = i(a^\dagger - a)$ then execute simple harmonic motion:
\begin{eqnarray}
\hat{X}(t_I) &=& \cos(\chi t_I \J_x)\hat{X}(0) + \sin(\chi t_I \J_x) \hat{Y}(0) \\
\hat{Y}(t_I) &=& \cos(\chi t_I \J_x)\hat{Y}(0) - \sin(\chi t_I \J_x) \hat{X}(0).
\end{eqnarray}

After time $t_I$, the light field is rapidly damped out.  If all the light is emptied from the cavity, then the integrated photocurrent for the measured quadrature ($\hat{X}$) is the true distribution for the quadrature\cite{Wiseman95,WisMil93}.  In terms of the Wigner function $W$, this is given by the marginal distribution.
\begin{equation}
p(x) = \frac{1}{4}\int^{\infty}_{-\infty}dy W(x,y),
\end{equation}
where $x = \alpha^* + \alpha$ and $y= i(\alpha^* - \alpha)$.

In general, the Wigner function will be a sum of gaussians, weighted by the atom number distribution of the condensate:
\begin{equation}
W(\alpha,\alpha^*) = \frac{2}{\pi}\sum^j_{m = -j} P_x(m) e^{-2|\alpha - \alpha_m'|^2}
\end{equation}
where $\alpha_m' = \alpha_0e^{-i\chi t_I m} = \frac{1}{2}(x(0) + iy(0))e^{-i\chi t_I m}$.  For convenience, we set the initial conditions of the light field such that $x(0)=0$.  This gives
\begin{equation}
W(x,y) = \frac{2}{\pi}\sum^j_{m = -j} P_x(m) e^{-\frac{1}{2}(x-y(0)\sin{\chi t_I m})^2 - \frac{1}{2}(y-y(0)\cos{\chi t_I m})^2}.
\end{equation} 
After integrating, the marginal distribution is
\begin{equation}
p(x) = \frac{1}{\sqrt{2\pi}} \sum^j_{m = -j} P_x(m) e^{-\frac{1}{2} (x-y(0) \sin{\chi t_I m})^2}.
\end{equation}
Thus each $m$ value is mapped onto a gaussian at position $y(0) \sin{\chi t_I m}$ with width equal to one.  This is illustrated in fig.(\ref{fig2}).  To be able to distinguish without ambiguity different $m$ values in the output, there should be at least 4 standard deviations between the means of adjacent gaussians.  The resultant condition on the atom-light coupling is then
\begin{equation}
\chi t_I > \frac{4}{|y(0)|} = \frac{2}{|\alpha_0|}
\end{equation}
 and 
\begin{equation}
\chi t_I \ll  \frac{1}{N}.
\end{equation}
If the fringes close to $m=0$ only are needed, then this last condition may be relaxed considerably to a condition which merely prevents aliasing: 
\begin{equation}
\chi t_I < \frac{\pi}{N}.
\end{equation}
Nevertheless, there is a limit on the size of the induced phase shift.

The analysis above assumes perfect detector efficiencies and an infinite time of integration so that all the light is removed from the cavity.  The results can be generalized to hold when this is not the case\cite{Wiseman94}.  For a detector efficiency of $\eta_\infty$ and a total integration time of $T$, then the distribution for the integrated photocurrent is
\begin{equation}
p(x) = \frac{1}{\sqrt{2\pi \eta}} \sum^j_{m = -j} P_x(m) e^{-\frac{1}{2} (x-\eta y(0) \sin{\chi t_I m})^2/\eta},
\end{equation}
where $\eta = \eta_\infty \left(1-e^{-\gamma T}\right)$ and $\gamma$ is the damping rate of the cavity.  The lower bound on the atom-light coupling becomes
\begin{equation}
\chi t_I > \frac{4}{\sqrt{\eta}|y(0)|} = \frac{2}{\sqrt{\eta}|\alpha_0|}.
\end{equation}
Thus the effects of detector inefficiencies and a low damping rate can be overcome by starting with a large coherent state amplitude in the cavity. For example, with detector efficiency of $\eta_\infty = .5$, $N_p = 10^7$ photons in the cavity and a measurement strength of $\chi = 10^{-2} s^{-1}$, then for $\gamma T \gg 1$, the lower limit on the interaction time is 
\begin{equation}
t_I > 90 ms
\end{equation}
This value of $\chi$ is calculated using a trap frequency of $\omega_0/2\pi = 32 Hz$, beam waist
$w=30\mu m$, light detuning $\delta/2\pi=100 MHz$, saturation intensity $I_s = 17 W/m^2$, optical 
frequency $\omega/2\pi = 3.8 \times 10^{14} Hz$, atomic linewidth
$\Gamma_a/2\pi = 10^7 Hz$ and
incident power $P=6 mW$, in a cavity $10cm$ long. 

\section{Phase errors and phase diffusion}

Interatomic collisions are necessary to produce the initial superposition of condensates in the first stage of the scheme, but their effect on subsequent stages of the measurement scheme is unwanted.  We have assumed that they can be suppressed using a Feshbach resonance.  However, it may be unfeasible to use this technique, so we now discuss briefly the effect of the nonlinearity on the different stages of the scheme.

In the second part of the measurement scheme, when the tunneling is turned off, the size of the self-interaction term may be comparable to that of the weak force term.  This will induce an extra self-phase change:
\begin{equation}
\phi = - |\kappa| j^2 \tau.
\end{equation}
However, since this self-phase change is the same for both components of the superposition, there is no net effect on the output probability distribution (eq.(\ref{prob})).

In the third stage, due to the constraints of the two-mode approximations, there are limits to the size of the interwell coupling.  If the atomic collisions are weak compared to the tunneling term, then the nonlinear term will cause a collapse (through dephasing) in any tunneling oscillations.  However, this should not occur before there is time for at least one quarter of an oscillation (a $\pi/2$ pulse) to occur.  If the self-interactions are stronger, then they will induce a nonlinear rotation around the $\J_x$ axis and a diffusion of the distribution on the Bloch sphere.  The effect of the extra rotation may be negated by adjusting the time of the pulse so that the final state lies in the $\J_y-\J_z$ plane.  The effect of the diffusion can not be so removed, and may wash out the interference fringes.  

Finally, there is last stage of the measurement when the condensate interacts with the cavity field.  If there is no tunneling, the collisions will have no direct effect on the state of the light field, since the $\kappa\J_x^2$ term in the hamiltonian doesn't effect the $x$-distribution.  
If the $\Omega$ is not exactly zero then, for a strong atom-light interaction, a back action may develop over time which will induce, through momentum fluctuations, tunneling\cite{CorMil98}.  This would directly affect the phase of the cavity field, and it also may open a way for the atom-atom interaction to have an effect.

The effect of the back action on the condensate may be seen in the master equation for the system in which the dynamics of the optical field has been adiabatically eliminated\cite{MilJacWal94}:
\begin{eqnarray}
\dot{\rho} &=& -i\Omega[\hat{J}_z,\rho]-i 2\kappa[\hat{J}_x^2,\rho] \nonumber \\
	   &+& i \chi |\alpha|^2 [\hat{J}_x, \rho] - \frac{2 \chi^2|\alpha|^2}{\gamma}[\hat{J}_x, [\hat{J}_x, \rho]].
\end{eqnarray}
The last term in this equation is the decoherence induced by the external coupling (to the cavity field).  Note that since it only involves $\J_x$ operators, if $\Omega$ is zero then the decoherence cannot affect the $x$-distribution and hence the induced light shift in the field.  

To avoid the waiting time involved in switching on the cavity field during the measurement process, it may be necessary to have the cavity on before the measurement begins.  This will mean that the decoherence is active during the detection stage of the scheme, and so a random phase will be imparted to the condensate superposition.  The double commutator in the master equation may be simulated by a stochastic term in the hamiltonian:
\begin{eqnarray}
\hat{H}_S = \frac{2\chi|\alpha|}{\sqrt{\gamma}}\frac{dW}{dt}\J_x,
\end{eqnarray}
where $dW$ is the Weiner increment.  The resultant phase change has a standard deviation that grows with the square root of time:
\begin{eqnarray}
\sigma_\phi(\tau) = \frac{2\chi|\alpha|}{\sqrt{\gamma}} j\sqrt{\tau} = \frac{2\chi|\alpha|}{\sqrt{\gamma \tau}\Delta}\ex{\phi(\tau)}.
\end{eqnarray}
Thus the relative error caused by this phase diffusion may be minimized by increasing the detection time $\tau$ or decreasing the strength of the interaction with the optical field.  For typical parameters (as used above), with $j = 50$ and $\tau = 160 ms$, the phase error is $\sigma_\phi = 0.08$ {\em rad}, which could be a major restriction on the sensitivity of the measurement.  Hence it may be better to switch on the cavity only when it is time to use the optical field.

\section{Mean Field Limit}

The scheme outlined above depended on starting in a state which was a quantum superposition of two condensates and on the resulting entanglement.  For comparison, we now present the mean field analogue to show what features of this scheme remain in the absence of quantum entanglement.

In the mean field limit, the system may be described by a Gross-Pitaevskii Equation\cite{YooNeg77,LifPit89,RupHolBur95} (GPE):
\begin{eqnarray} 
i\hbar \dot{\Phi}(x,t) &=& \left(-\frac{\hbar^2}{2 M} \frac{\partial^2}{\partial x^2} + Rx + V(x) +U_0|\Phi(x,t)|^2\right)\Phi(x,t), \label{GPE}
\end{eqnarray}
where $M$ is the atomic mass, the constant $R$ is the gradient of the single particle potential due to the force and $U_0$ is the strength of the interparticle interactions.  Suppose that the interatomic collisions are negligible.  Then, when the overlap between the wells is small, we may expand the mean field in terms of the local wavefunctions of each well:
\begin{eqnarray} 
\Phi(x,t) = b_1(t)u_1(x) + b_2(t)u_2(x)
\end{eqnarray}
where 
\begin{eqnarray}
u_j(x) = e^{-iE_0/\hbar}\frac{1}{(2\pi r_0)^{\frac{1}{4}}} e^{(x-(-1)^jx_0)^2/4r_0^2}\:,
\:r_0=\sqrt{\frac{\hbar}{2M\omega_0}}  
\end{eqnarray}
and
\begin{eqnarray}
b_j(t) = \int u_j^*(x)\Phi(x,t)dx.
\end{eqnarray}
The ground state energy of each of the local modes is $E_0$.
From the GPE (eq.(\ref{GPE} ), the resultant equations of motion for the $b_j(t)$s are 
\begin{eqnarray}
\dot{b}_j(t) &=& \frac{-(-1)^jiRx_0}{\hbar}b_j(t) + \frac{i\Omega}{2}b_{3-j}(t) \\
\end{eqnarray}

As before, to perform the weak force measurement, we allow the force to act for a time $\tau$ and then the tunneling for a time $t_{\pi/2} = \frac{\pi}{2\Omega}$:
\begin{eqnarray}
b_j(\tau+t_{\pi/2}) &=& e^{-(-1)^jiRx_0\tau/\hbar}b_j(0)\cos{\frac{\Omega t_{\pi/2}}{2}} + ie^{-(-1)^{j+1}iRx_0\tau/\hbar}b_{3-j}(0)\sin{\frac{\Omega t_{\pi/2}}{2}}\\
\end{eqnarray}
Suppose we start off with an equal occupation in each well, such that $b_1(0)=b_2(0)=\sqrt{N/2}$.  Then the mean population difference is shifted by the presence of the weak force:
\begin{eqnarray}
\ex{m} &=& \frac{1}{2}\left( |b_2(\tau+t_{\pi/2})|^2 - |b_1(\tau+t_{\pi/2})|^2 \right) \\
 &=& -\frac{N}{2}\sin{\frac{2Rx_0\tau}{\hbar}}\\
 &=& -\frac{N}{2}\sin{\Delta\tau}. \label{mfr}
\end{eqnarray}

This did not occur in the previous quantum treatment, in which $\ex{m} \equiv 0$.  Even when the system is simply in a number state (not a superposition) with an equal number of atoms in each well, i.e. $\ket{j,0}_x$, the mean population difference is unaffected by the presence of the force.  This is a demonstration of the fact that, as a classical treatment, the mean field situation cannot be regarded as the large number limit of a quantum number state.  In quantum optics, it is the minimum uncertainty coherent state $\ket{\alpha}$ which is most like a mean field with amplitude $\alpha$.  The analogue in this case is the atomic coherent state, or Bloch state:

\begin{eqnarray}
\ket{\beta} = \sum_m \left( \begin{array}{c} 2j \\ m+j \end{array}\right) \frac{\beta^{m+j}}{(1+|\beta|^2)^j}\ket{j,m}_z,
\end{eqnarray}
where $\beta$ can be described in terms of the angular coordinates of a point on the sphere $\beta = \tan\theta e^{i\psi}$.

Consider the state given by $\beta =0$, which is symmetric with respect to the two wells.  This state is also the ground state $\ket{j,-j}_z$ of the system when $\Omega \gg |\kappa|N$, in other words, a state in which the coherence between the two condensates is well established through tunneling.  If the system begins in this state, then after the measurement procedure, the difference in occupation between the two wells is as given above in the mean field approach (eq.(\ref{mfr})).

It may seem better to use this Bloch state as the initial state, since it may be easier to generate than the superposition state previously used (Eq.(\ref{gnd}) and the sign of the weak force may be determined from the measurement of $\ex{m}$.  However, the size of the induced phase change given in Eq.(\ref{mfr})is {\em not} amplified by $j$.  In other words, the macroscopic occupation of a single condensed state in not being fully utilized.

A comparison of the relative uncertainty in either case clearly demonstrates this point.  In the first case, where the superposition state is used, the phase is inferred by the proportion of detection events falling on odd fringes (Eq.(\ref{odd})).  The uncertainty in this binomial distribution with probability $P= \sin^2{\Delta j\tau}$ is 
\begin{equation}
\delta P = \sqrt{\frac{P(1-P)}{N_P}},
\end{equation} 
where $N_P$ is the number of detection events.  The relative uncertainty in the phase is thus
\begin{eqnarray}
\frac{\delta \phi}{\phi} = \frac{1}{\phi} \left| \frac{dP}{d\phi}\right|^{-1} \delta P = \frac{1}{\Delta \tau N \sqrt{N_P}}. 
\end{eqnarray}
When the initial state is the Bloch coherent state $\ket{j,-j}$, the uncertainty in the mean of the distribution is 
\begin{equation}
\delta \ex{m} = \sqrt{\frac{\ex{m^2} - \ex{m}^2}{N_P}}.
\end{equation} 
This gives a relative uncertainty in $\phi$ of
\begin{eqnarray}
\frac{\delta \phi}{\phi} = \frac{1}{\phi} \left| \frac{d\ex{m}}{d\phi}\right|^{-1} \delta \ex{m} = \frac{1}{\Delta \tau \sqrt{N N_P}}. 
\end{eqnarray}

Thus the precision of the measurement grows in proportion to the number of condensed atoms when the entangled state (Eq.(\ref{gnd})) is used, but only as the square root of the number of atoms when the coherent state $\ket{j,-j}$ is used.  This demonstrates the advantage of using quantum entanglement of two macroscopically distinct states in order to make a highly sensitive force detector.
\section{Conclusions}
If the weak force in question is gravity, then, with the parameters quoted above, the size of the induced phase shift is
\begin{equation}
\frac{\phi}{g} = \frac{NM\tau x_0}{\hbar} \simeq 2.4\times 10^4  rad{\rm \:per \:}g, 
\end{equation} 
where $x_0 = 15 \mu m$ and $M = 10^{-26} kg$ (for lithium).  While this is small compared to the phase shift which atom interferometric techniques\cite{KasChu91,PetChuChu97,YouKasChu97} can obtain ($\simeq 3 \times 10^{6}rad$ per $g$), improvements can be made.  The size of the phase shift may be increased using more atoms, separating the wells further or by allowing a longer time $\tau$ for the interaction.  The number of atoms could be increased by up to $10$ times without invalidating the two mode approximation.  The two wells need to remain close during the preparation of the initial state and during tunneling, but could be separated and brought together againg during the weak force interaction.  The time of the interaction is limited by mechanical vibration.

Hence using a double condensate in a scheme to make sensitive measurements, such as that presented here, may be feasible.  The experimental techniques currently being developed to produce and manipulate BECs may allow such a scheme to be realized in the near future.

\begin{figure}
\caption{Schematic outline of proposed measurement scheme.  The stages are: (a) preparation of the superposition state, (b) interaction with the weak force for time $\tau$, (c) a $\pi/2$ rotation caused by tunneling, and (d) a homodyne measurement of the number of atoms in one well.}
\label{fig1}
\end{figure}

\begin{figure}
\caption{The probability distribution resulting from the homodyne measurement.  The fringes corresponding to adjacent $m$ values can be distinguished if the coherent amplitude of the light field $\alpha_0$ is large enough.}
\label{fig2}
\end{figure}


\begin{references}
\bibitem{Anderson95}	M. H. Anderson, J. R. Ensher, M. R. Matthews, C. E. 
Wieman,
                        and E.A.Cornell,
                        {\sl Science} {\bf 269}, 198 (1995).
%
\bibitem{Hulet95}   	C. C. Bradley, C. A. Sackett, J. J. Tollett, and R. G.
Hulet,
			            {\sl Phys. Rev. Lett.} {\bf 75}, 1687 
(1995).
%
\bibitem{Ketterle95}	K. B. Davies, M. -O. Mewes, M. R. Andrews, N. J. van
                        Druten,
		            	D. S. Durfee, D. M. Kurn, and W. Ketterle,
  			            {\sl Phys. Rev. Lett.} {\bf 75}, 3969 
(1995).
%
\bibitem{BEC97} Reports on more recent experiments can be found on 
http://amo.phy.gasou.edu/bec.html
\bibitem{ScienceNews} Report in ``News of the Week'', {\sl Science} {\bf 281}, 501 (1998).
\bibitem{HalMatetal98} D. S. Hall, M. R. Matthews, J. R. Ensher, C. E. Wieman and E. A. Cornell, {\sl Phys. Rev. Lett.} {\bf 81}, 1539 (1998). 

\bibitem{HalMatWieCor98} D. S. Hall, M. R. Matthews, C. E. Wieman and E. A. Cornell, {\sl Phys. Rev. Lett.} {\bf 81}, 1543 (1998).
\bibitem{HueMacetal97} S. F. Huelga, C. Macchiavello, T. Pellizzari, A. K. Ekert, M. B. Plenio and J. I. Cirac {\sl Phys. Rev. Lett.} {\bf 79} 3865 (1997).

\bibitem{MilCorWriWal97} G.J. Milburn, J.F. Corney, E.M. Wright and D.F. Walls,
		         {\sl Phys. Rev. A} {\bf 55}, 4318 (1997).
\bibitem{CorMil98} J.F. Corney, G.J. Milburn, {\sl Phys. Rev. A} {\bf 58}, 2399 				(1998).
\bibitem{HarWisMil98} 	D. J. Harris, H. M. Wiseman and G. J. Milburn, {\sl unpublished}, (1998).
\bibitem{SteCol98}	M. J. Steel and M. J. Collett, {\sl Phys. Rev. A} {\bf 57}, 2920 (1998).
\bibitem{CirLewMolZol98} J. I. Cirac, M. Lewenstein, K. M\"olmer and P. Zoller, {\sl Phys. Rev. A} {\bf 57}, 1208 (1998).

\bibitem{InoAndetal98} S. Inouye, M. R. Andrews, J. Stenger, H.-J. Miessner, D. M. Stamper-Kurn and W. Ketterle, {\sl Nature} {\bf 392}, 151 (1998).
\
\bibitem{Wiseman95}     H. M. Wiseman, {\sl Quantum and Semiclassical Optics}, 					{\bf 7}, 569 (1995) 
\bibitem{WisMil93}	H. M. Wiseman and G. J. Milburn, {\sl Phys. Rev. A} 
			{\bf 47}, 642 (1993).
\bibitem{Wiseman94}	H. M. Wiseman, {\sl PhD Thesis}, The University of Queensland, 1994, p63.
\bibitem{MilJacWal94}	G. J. Milburn, K. Jacobs and D. F. Walls, 
			{\sl Phys. Rev. A} {\bf 50}, 5256 (1994).
\bibitem{YooNeg77}      B. Yoon and J. W. Negele,
                        {\sl Phys. Rev. A} {\bf 16}, 1451 (1977).
\bibitem{LifPit89}      E. M. Lifshitz and L. P. Pitaevskii, 
                        {\it Statistical Physics, Part 2} 
                        (Pergamon Press, Oxford, 1989)pp. 85-118.  
\bibitem{RupHolBur95}   P. A. Ruprecht, M. J. Holland, K. Burnett,
                        and M. Edwards,
                        {\sl Phys. Rev. A} {\bf 51}, 4704 (1995).

\bibitem{KasChu91}	Mark Kasevich and Steven Chu, {sl Phys. Rev. Lett.} {\bf 67}, 181 (1991).
\bibitem{PetChuChu97} 	Achim Peters, Keng Yeow Chung and Steven Chu,``Atom interferometry applied: Measuring gravity with ultra high precision'', {\sl IV Workshop on Optics and Interferometry with Atoms}, St. John's College Oxford, 21st-23rd July 1997.
\bibitem{YouKasChu97}   Brenton Young, Mark Kasevich and Steven Chu, ``Precision Atom Interferometry with Light Pulses'', {\sl Atom Inteferometry}, ed. Paul R. Berman, Academic Press, 1997.

\end{references}
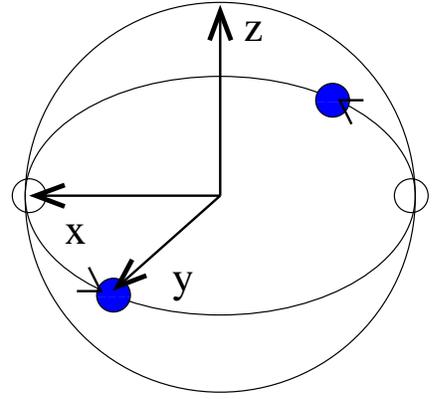
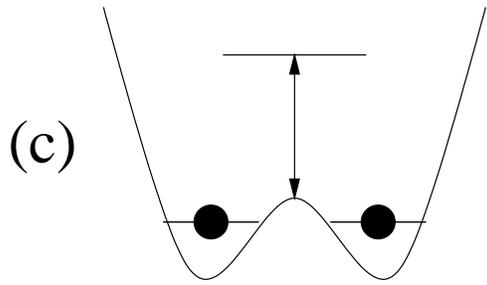
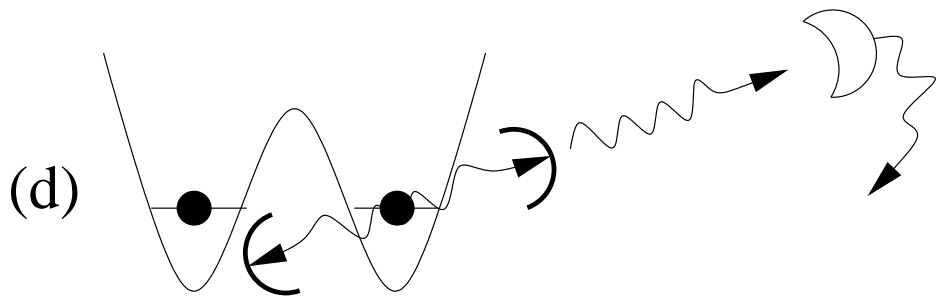
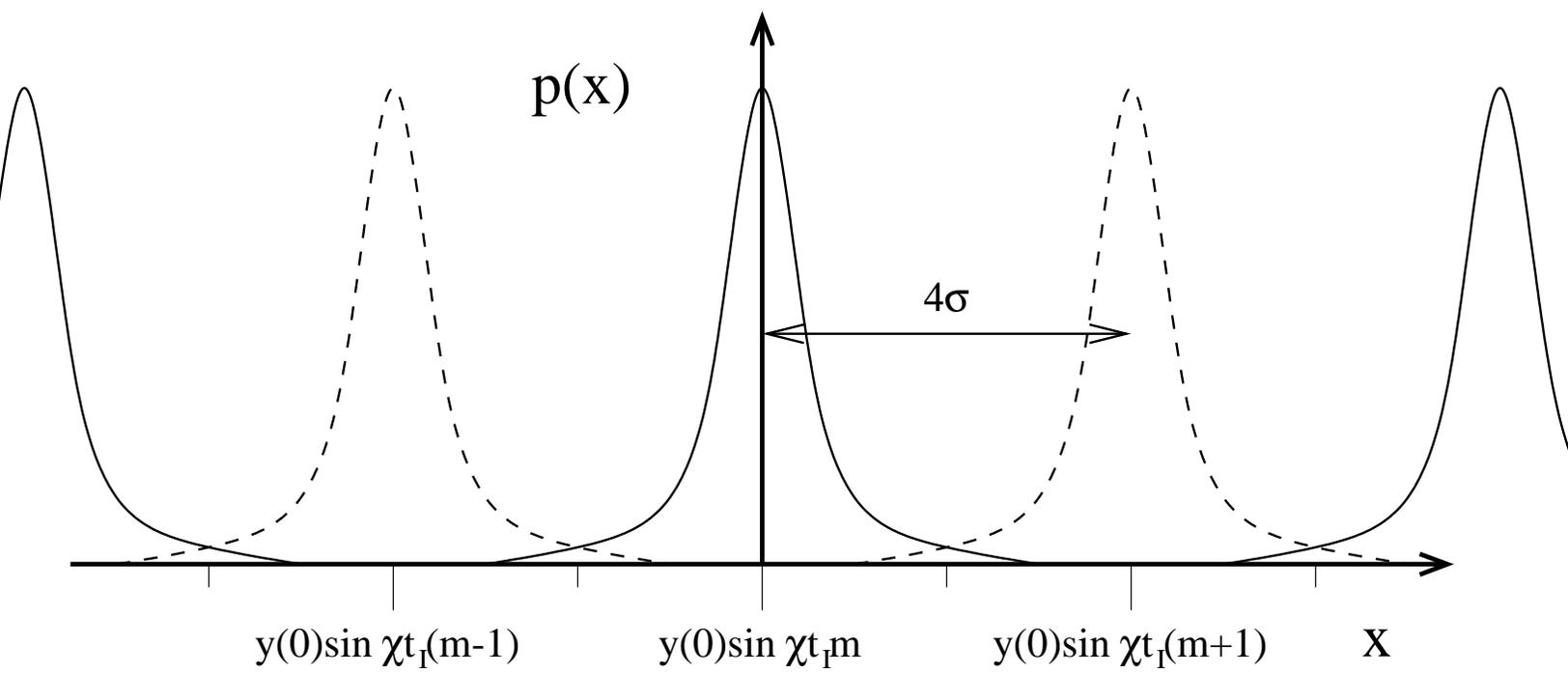
\end{document}